# Megadiff: A Dataset of 600k Java Source Code Changes Categorized by Diff Size

Martin Monperrus    Matias Martinez    He Ye
Fernanda Madeiral    Thomas Durieux    Zhongxing Yu

**Abstract**

This paper presents Megadiff, a dataset of source code diffs. It focuses on Java, with strict inclusion criteria based on commit message and diff size. Megadiff contains 663 029 Java diffs that can be used for research on commit comprehension, fault localization, automated program repair, and machine learning on code changes.

## 1  Introduction

Researchers have analyzed *source code changes* to create knowledge on the evolution of software systems and, consequently, to create new techniques and tools to support software development tasks. For instance, Le et al. [9] prioritize mutation operators for program repair based on their prevalence in source code diffs. Long and Rinard [13] train the patch generation system by learning human diffs obtained from open-source software repositories.

Researchers tend to create ad-hoc datasets of source code diffs for the tasks at hand. This poses several problems with respect to scientific reproducibility. First, the dataset of diffs may not be made publicly available. This leads to one approach cannot be compared with another based on the same dataset. Second, the diff inclusion criterion may not be documented in the paper, due to focusing on another problem. Different criterion leads to the collected diffs in different quality, which may further impact the experimental results. Third, the dataset of diffs may be shared only in a processed format, not in raw format, making reuse impossible out of the scope of the already done prepossessing.

To solve this problem, this paper proposes Megadiff, a novel dataset of source code diffs focusing on Java. Here, a diff is defined as a set of line-based changes made to a source code file. Megadiff concentrates on small changes made to Java files. Megadiff solves the aforementioned problems with the following characteristics: 1) it is publicly available on GitHub and Zenodo; 2) the inclusion criteria are systematic; 3) the dataset contains the raw diff, with all lines of the changed files, with no exception. 4) Megadiff is well categorized by diff size, ranging from 1 to 40 lines of code changes, which provides wider choices for research experiments (in contrast to diffs with specific lines of code change).

While it is straightforward to take any GitHub repository to study software evolution, it is difficult to do so at scale. At full scale (with a large number of repositories and a large number of commits), it is time-consuming to collect, filter and analyze commits.

The design of Megadiff hides this complexity for the researchers, and the scale of Megadiff results from weeks of computation for data collection and filtering.

**To sum up, our contribution is Megadiff, a dataset of source code diffs in the Java programming language. To our knowledge, this is the largest dataset of Java diffs ever.**

## 2 Collection Process

In order to create MegaDiff, we use the BOA platform [4], a domain-specific language and infrastructure that eases mining software repositories. BOA platform allows easy execution queries on several datasets. The queries allow, for example, to look for specific AST constructions or finding commits that have a specific commit message. In our case, we use the biggest Boa dataset: "2015 September/GitHub". It is composed of more than 380,000 code repositories and 23 million commits.

**Commit and project inclusion criterion** We use the following Boa query to collect the commit and the projects.

```
out: output collection[string] of string;
p: Project = input;

exists (l: int; match('java', lowercase(p.programming_languages[l])))
foreach (i: int; def(p.code_repositories[i]))
        foreach (j: int; def(p.code_repositories[i].revisions[j]))
            if (isfixingrevision(p.code_repositories[i].revisions[j].log))
                out[p.code_repositories[i].revisions[j].id]
                    << p.code_repositories[i].url;
```

This query iterates on all commits of all Java projects from the dataset. If the commit is a fixing commit (see http://boa.cs.iastate.edu/docs/dsl-functions.php#isfixingrevision), we collect the commit SHA and the URLs of the project.

We execute this query on April 2nd, 2019 which resulted in 4 590 405 unique commit identifiers. Those commits are spread over 101 472 unique GitHub repositories. On April 20, 2019, we have attempted to clone them all and found that 14 816 are no longer publicly available. At this date, there were 3 934 926 commit identifiers from BOA that are in public and available repositories.

**Commit filtering** The next step was to filter the commits. We select commits which: 1) contain changes in at least one Java source file, 2) contain less than n changed lines of code in *.java files (as defined as the sum of added and removed lines, as given by `git diff`). We set n = 40 given out storage budget.

**Data preparation** For each commit, we extract the diff with full context and store it in a single file. This means that the source code around and involved in the diff is raw, unmodified, unprocessed Java source code.

*Finally, Megadiff is composed of 663 029 commits done on Java source code.*

# 3  Content

**Repository Content** Megadiff is available as a Git repository at `https://github.com/monperrus/megadiff` and preserved on `https://doi.org/10.5281/zenodo.5013515`. The size of Megadiff is 4.6 GB (considering all diff files compressed).

**Folder content** The diffs are sorted in the folders. Each folder refers to the number of changed lines (the sum of additions and deletions) in the diff. For instance, folder 2 contains diffs with a) 2 added lines;b) 2 deleted lines; c) 1 addition and one deletion.

**File content** There is one diff file per commit. The diff can affect different files. The diffs follow the unified diff format. For sake of storage, all diff files are compressed with the XZ utility. The diff files have the following characteristics:

- The diff file name is the commit SHA1. Hence, it is possible to use GitHub to search for finding the origin of the diff as follows:
  `https://github.com/search?q=acbfff0a50baaceae0ad1e1160fffa693b50c994&type=Commits`.

- A diff is composed of all Java file changes (a commit can change several Java files at once).

- The diffs are given with full context, meaning that both the full files before and after the change are present.

- The diffs are unfiltered: most of them contain code changes, but some of them only contain formatting or documentation changes.

# 4  Applications

Megadiff can be used in several research topics for researchers aiming at different goals. One major application is the study of how developers evolve software systems. More concretely, source code diffs could be analyzed in terms of maintenance activities (corrective: bug fixing; perfective: system improvements; adaptive: new feature introduction), like done by Levin and Yehudai [11].

Furthermore, the diffs related to particular maintenance activities could be used for research in specific research fields. For example, the diffs related to bug fixing could be used for understanding how developers fix bugs [22, 20, 14].

On a more technical side, diffs could even be used to purely evaluate tools. For instance, Megadiff could be used to evaluate specialized tools that extract features from diffs, e.g. [15]. Note that there exists a usage of Megadiff in this direction: the diffs are used to evaluate an approach for hyperparameter optimization of AST differencing [17].

Other than understanding source code changes and using them for evaluating approaches and tools, Megadiff can also be used to feed systems. For instance, the diffs can be used for the training of supervised-machine-learning-based program repair tools, e.g., [1]. In that direction, Yu et colleagues use Megadiff to learn source code transformation and apply the learned model to predict transformation that repair bugs.

Table 1: Datasets of Java source code diffs.

| Dataset | # Diffs |
|---|---:|
| CVS-Vintage [18] | 89,993 |
| HDRepair [9] | 16,450 |
| CodRep [2] | 58,069 |
| Fixminer [8] | 11,416 |
| IntroClassJava [3] | 297 |
| ManySStuBs4J [7] | 153,652 |
| Megadiff (this paper) | 663 029 |

# 5 Related datasets

There are several datasets of source code diffs. In this section, we discuss the ones focused on Java, summarized in Table 1.

Martinez and Monperrus' CVS-Vontage [18] is a dataset of 89 993 CVS transactions in Java. The limitations of this dataset are that 1) it is only composed of old Java software (i.e., written in early versions of the language, with the programming practices of that time) 2) CVS transactions are file-based, there is no concept of commit aggregating changes over different files in CVS.

Le and colleagues [9] have collected 3202 Java diffs and shared them in an open-source repository. Liu et al. [12] consider 16 450 commits over 6 Java projects, shared in a replication package. Koyuncu et al. [8] have collected and used 11,416 Java patches. The scale of those datasets is relatively small. CodRep is a dataset of diffs [2], which only focuses on one-liners (those diffs that change a single line). This prevents CodRep to be representative of larger changes. IntroClassJava from Durieux and Monperrus [3] is the Java version of IntroClass from Le Goues and colleagues [10], which contains 297 small buggy programs (5-20 lines) written by students. The dataset also includes the reference version (correct) of such buggy programs.

There are other related datasets, such as Defects4J [5], Bugs.jar [19], and Bears [16]. These datasets are publicly available and have been useful for research, mainly in automated program repair. While they make the source code diff of bug fixes available, they are small in comparison with Megadiff due to their different nature of containing reproducible bugs.

There are many other papers studying diffs in the context of mining software repositories, but few have consolidated and shared a dataset. For instance, Zhong and Su [22] study 15 119 commits in Java software, but they have only shared the metadata of the commits, not the actual source code of the diffs. Soto and colleagues [21] have used the BOA platform to study Java software evolution, but have not collected the diffs. Karampatsis and Sutton [6] study the changes involved in bug fixes that modify one single statement. For that, they created a dataset named ManySStuBs4J [7], composed of 153,652 single statement bug-fix changes mined from 1,000 popular open-source Java projects. As a difference, Megadiff includes commits that affect more than one statement.

# 6 Conclusion

Megadiff is a dataset of 663 029 diffs of Java source code, taken from open-source projects. To our knowledge, this is the largest dataset of Java source code diffs to date. We envision that Megadiff will provide useful data for future research about software evolution and machine learning on code.